# From Keywords to Clusters: AI-Driven Analysis of YouTube Comments to Reveal Election Issue Salience in 2024

**Authors:**

Raisa M. Simoes, Ph.D.: Adjunct Lecturer Georgia Tech-Europe, Sam Nunn School of International Affairs

Timoteo Kelly, MSHI: PhD Student, NSF CyberCorps Scholar, University of Missouri Institute for Data Science and Informatics

Eduardo J. Simoes, MD, MSc, MPH: Distinguished Professor and Director of Center for Medical Epidemiology and Population Health, University of Missouri School of Medicine

Praveen Rao, Ph.D.: Associate Professor, Dept. Of Electrical Engineering & Computer Science, University of Missouri

Corresponding author: Raisa M. Simoes; email: simoes.raisa@gmail.com

**Abstract:**

This paper aims to explore two competing data science methodologies to attempt answering the question, "Which issues contributed most to voters' choice in the 2024 presidential election?" The methodologies involve novel empirical evidence driven by artificial intelligence (AI) techniques. By using two distinct methods based on natural language processing and clustering analysis to mine over eight thousand user comments on election-related YouTube videos from one right leaning journal, Wall Street Journal, and one left leaning journal, New York Times, during pre-election week, we quantify the frequency of selected issue areas among user comments to infer which issues were most salient to potential voters in the seven days preceding the November 5th election. Empirically, we primarily demonstrate that *immigration* and *democracy* were the most frequently and consistently invoked issues in user comments on the analyzed YouTube videos, followed by the issue of identity politics, while inflation was significantly less frequently referenced. These results corroborate certain findings of post-election surveys but also refute the supposed importance of inflation as an election issue. This indicates that variations on opinion mining, with their analysis of raw user data online, can be more revealing than polling and surveys for analyzing election outcomes.

**Keywords**: artificial intelligence; opinion mining; clustering; vote choice; cleavages

1. **Introduction**

The Democrats lost both houses of Congress and the Presidency to Republicans in the 2024 election, with former president Donald Trump winning all seven swing states and the national popular vote, despite most pre-election polls giving Vice President Kamala Harris and President Trump a roughly equal chance of winning. Most post-election punditry and analysis in the legacy press and alternative media has attributed the Democrats' large loss to two main issues – inflation [59] and immigration [30] – as well as to structural conditions that do not favor incumbent parties [5, 23]. However, a growing contingent of analysts has also attributed the election outcome to the Democratic party's association with cultural issues purportedly distant from the median voter's preferences, such as those alternatively aggregated under the concept of "identity" or "woke" politics [54, 56].

To this point, three post-election studies illustrate how voters associated Democrats with left-of-center ideas that were ostensibly distant from most voters' priorities. Survey research from the think tank Third Way demonstrates that Democrats, and thus Kamala Harris, were largely perceived as "too liberal" [15], while a study from More In Common polling over 5,000 Americans concluded that while inflation was the top concern for every major demographic group across both parties, Americans misperceived LGBT/transgender policies as the top policy priority for Democrats [37]. In another post-election poll, Blueprint concluded that behind inflation and immigration, the third reason voters cited for not voting for Harris was her perceived greater focus on "cultural issues like transgender issues rather than helping the middle class" [8]. For swing voters in the Blueprint poll, this perceived outsized focus on cultural issues was the top reason for not voting for Harris, over inflation and immigration.

Despite the post-election survey and polling data, opposing factions within the Democratic party point to different factors in tipping the election to Trump [19]. There is little consensus on the reasons for Trump's convincing victory, and given the recent conclusion of the election, there is not yet scholarly literature available to begin establishing a plausible theory that explains the 2024 results. The aim of this article is to establish novel empirical evidence showing which issue(s) contributed to the outcome of the 2024 election by using artificial intelligence (AI) techniques to mine user comments on YouTube videos from mainstream media outlets and quantify the prevalence of certain issue areas among user comments. Taking into consideration post-election survey and polling data, as well as established research in political science demonstrating shifts in industrialized societies towards the salience of non-economic and cultural issues, we hypothesize that cultural or identity-related issues contributed at least as much, if not more, to Trump's victory as inflation or immigration. We operationalize this by developing a unique AI-based methodology to identify which issues are most frequently discussed by users of online platforms, these users representing potential voters. Additionally, we propose that this AI-driven variation of opinion mining provides an empirical alternative to polling as a method for forecasting election results.

## 2. Theoretical and methodological framework for examining electoral outcomes

### 2.1. Shifting sociopolitical cleavages: the growing cultural divide

Explanations for voter behavior are subjects of great debate in political science. Decades ago, the seminal work of Campbell [10] on the American voter highlighted the importance of party identification in determining vote choice. Since its publication, party identification and general political participation have declined across industrialized societies [18], though recent work by Abramowitz and Webster [2] on increasing political polarization and in particular "negative partisanship" or "negative affect toward the opposing party" [2, p. 17] may bring the effects of party identification back into focus. To this point, the study by Garzia and Ferreira da Silva [20] of the 2020 election shows that negative affective polarization does translate into "negative voting" *against* a candidate. One of the more persistent theories of voter decision making is that voters are most concerned with economic well-being [32], which underlies 1992 Democratic campaign advisor James Carville's famous phrase that "it's the economy, stupid."

Our study seeks specifically to uncover the *issues* that drove vote choice in 2024, and we posit that inflation was not the single key issue driving voters' preferences. In this perspective, the landmark work of Ronald Inglehart on the "silent revolution" in political culture in industrialized societies is particularly informative [27]. Inglehart [28] hypothesized how an intergenerational value shift was occurring, in which younger cohorts born into greater wealth and economic security than previous generations were increasingly animated by post-materialist values regarding "autonomy and self-expression," [28, p. 130] such as gender roles and sexual mores that supplanted traditional cultural norms, and less by material concerns like economic and physical security. This research suggested that the traditional class cleavage of western societies was diminishing in comparison to cultural cleavages.

Though the post-materialist thesis was not necessarily a theory of voter behavior, it can help explain the general shift in political culture "from survival values to self-expression values" [28, p. 131] reflected in the issue areas among the main concerns of voters today. Indeed, the post-materialist framework helps explain the generalization of socially liberal or progressive values across many western societies, exemplified by increasing acceptance of gender equality, gay marriage, racial and ethnic diversity, etc. But Inglehart's more recent collaboration with Pippa Norris seeks to explain how this shift towards socially liberal post-material values has also engendered a "cultural backlash" [39]. As value changes across the western world have entrenched cultural cleavages, the reaction from social conservatives has had political consequences[1]. In addition, Mounk [38] demonstrates the growing prevalence of what he calls the "identity synthesis," in which group identity characteristics such as race and gender became the core objects of political action, particularly on the left. Pointing to the fall of Soviet communism as a turning point, Mounk notes that "the center of gravity on the left swung from class and economics to culture and identity" [38, p. 63]. The oft cited, though unscientific and ill defined, concept of "wokeness" falls under this umbrella concept of identity politics or identity synthesis.

It is also important to underscore how exceptional circumstances post-2020 generated two other non-economic issues that we believe could also have been particularly relevant for voters: public health and democracy. Starting in 2020, the world went through the largest infectious disease epidemic in 100 years. The government reaction to the COVID-19 pandemic in the United States, including vaccine mandates, lockdowns and subsequent job losses, school closures, and allegations of government censorship of pandemic-related information, put public health issues and institutions on the political map in an unprecedented manner. Most pertinently for an election year, a study from Pew Research Center [46] demonstrates how Americans' reactions to pandemic-related topics (vaccines, masks, etc.) fell largely along partisan lines, with, for example, 79% of Democrats responding that public health officials' response to COVID-19 was good or excellent, as opposed to only 35% of Republicans.  This politicized public response to the pandemic provided the backdrop for the Make America Health Again (MAHA) movement led by Robert F. Kennedy Jr, his presidential candidacy and eventual withdrawal and endorsement of Donald Trump. We therefore include public health and MAHA as another relevant non-economic issue to consider in online discussion.

---

[1] Inglehart and Norris argue that cultural backlash has contributed to the increasing representation of "authoritarian populist" parties or leaders across Europe and the United States in particular. The characterization of the *ideological* nature of the issues or parties that gained prominence with the substitution of economic cleavages with cultural cleavages is not within the scope of this paper. And the extent to which the Republican party led by Donald Trump, for example, represents authoritarian as opposed to democratic values is a subject of debate that we will not address here. Our study uses the cultural backlash to post-materialism only as a framework for understanding the salience of cultural *issues* in this year's election, as opposed to strictly materialist issues.

Additionally, democracy is another non-economic issue that we believe may have been salient for voters in 2024. This is due to the circumstances surrounding the 2020 election, including President Trump's indictment over alleged efforts to overturn the election results and the subsequent appointment of a special counsel to investigate the allegations, and the January 6$^{th}$, 2021 attack on the Capitol during the 2020 election certification. Indeed, President Biden's assertion in January 2024 that "democracy is on the ballot" in the 2024 election assured that the protection of democracy would be central to his campaign [47]. Vice-president Harris maintained this message on the campaign trail as well [3].

Overall, we argue that this reorientation of modern politics around cultural and identity-driven cleavages provides a framework for understanding the outcome of the 2024 election. This is not to say that issues of economic security, such as inflation and employment, play no role in voter behavior. However, there is insufficient empirical analysis of the explanatory weight of modern cultural and identity-focused issues, such as transgender rights/gender identity or critical race theory, as well as other non-economic issues like public health and democracy, in this year's election.

## 2.2. Opinion mining and its variations as relevant methods for uncovering salient issue areas

We will explore the relative salience of different issues prior to the 2024 election using novel methodology, by mining user comments on election-related YouTube videos from two news outlets. Our method is based in the logic of opinion mining, which, according to Sobkowicz et al. [50] includes social media content analysis and methods for modeling, simulating and predicting opinion formation "to create a knowledgebase containing online opinions in a more structured and explicit form" [50, p. 471]. In contrast to polling or survey methods, which contain risks of responder bias or dishonesty [13, 14] as well as bias built into the polls themselves [41, 49], the potential anonymity of online commentary can decrease the likelihood of self-censorship [52]. Mining comments for salient issues thus provides raw thinking from potential voters, allowing us to interpret which issues were most important to potential voters before the election. Analyzing this data over time could open the possibility of a new predictive method that is an alternative to polling in forecasting election results. Though we do not employ sentiment analysis, and thus no positive or negative sentiment can be attributed to the mention of specific issues, we do use novel AI methods to draw from the rich social media content. This variation on opinion mining allows for a preliminary interpretation of the issues that may have driven voters in 2024.

Unlike classic opinion mining, which often centers on sentiment analysis [34, 42], our approach focuses on identifying and interpreting issue salience for the 2024 presidential election through semantic clustering and AI-assisted labeling. This method allows us to classify discourse around our five predefined issue areas while allowing new themes to emerge. It allows us a more flexible and scalable variation on traditional opinion mining, better suited for capturing the complexity of social media comments [1, 17].

### 3. Research design

**3.1 Research question and Hypothesis:**

Which issues contributed to voters' choice of Donald Trump in the 2024 presidential election? Our hypothesis is that inflation and immigration were not the only key issues driving voters, and that cultural issues and non-economic issues related to identity politics, democracy and public health were at least as important or more important than inflation and immigration. The hypothesis arose from observation of hundreds of media articles throughout 2024 (mainstream press, independent media, podcasts, etc.) and particularly reader and listener comments on these articles and podcasts. By focusing not on polls or surveys but on analysis of these comments, this research provides a novel methodological approach for operationalizing the study of voters' choices in elections.

**3.2. Methodology and Data model:**

**Figure 1. Methodology flow chart**.

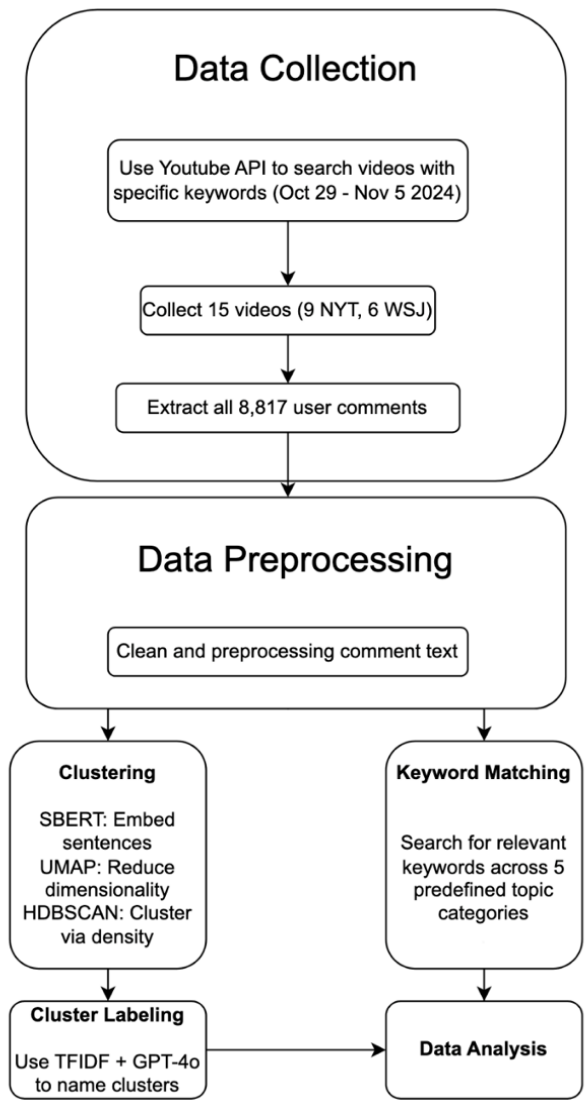

To quantify which issues contributed to voter choice, we used YouTube API package in Python to scrape comments from viewers of election-related videos on the YouTube channels of the Wall Street Journal (WSJ) and the New York Times (NYT). These two media outlets were selected for their relative representativeness of the political spectrum, research having shown that the WSJ tends to reflect a center-right readership and partisan leaning and the NYT a center-left readership and partisan leaning [16, 45].

The flowchart of the process is illustrated in Figure 1. We used the YouTube API to search for YouTube videos between October 29 2024 and November 5 2024 from the NYT and WSJ channels that fell under the keywords 'Donald Trump', 'Kamala Harris', 'Joe Biden', 'Tim Walz', 'Robert F. Kennedy Jr', 'RFK Jr', 'JD Vance', 'Democrats', 'Democratic Party', 'Republicans', 'Republican Party', 'DNC', 'RNC', 'Democratic National Convention', 'Republican National Convention', 'Election', 'Presidential Election 2024', and '2024 Election'. Specifically, the aforementioned keywords were searched under the YouTube video's tags, descriptions and titles. The keyword list was designed to capture a broad range of content relevant to the 2024 election cycle. This search produced fifteen YouTube videos, six from the WSJ and nine from NYT. Four of the videos were published on election day, two from each respective channel.

Once we obtained this list, we scraped and collected all user comments from every video on the list (with 15 videos) using YouTube API once again. This resulted in a sample of 8,817 comments. Once the comments were cleaned to remove duplicates, we were left with a sample size of 7,809, with 4,157 top comments and 3,652 replies. We then searched for specific keywords (Table 1) in the comments representing both economic and cultural or identity-related issues. Before running the search, we selected five main issue areas that were most relevant prior to the election: immigration, inflation, democracy, identity politics, and public health. We identified several keywords that reflected each issue area, based on literature review and observation of a vast range of media content. With these parameters, we conducted a search through the user comments with the keywords relating to our five categories. See Table 1 for the complete list of issue areas and their respective keywords. For example, in order to aggregate comments pertaining to the issue area "identity politics," we identified the following keywords: identity politics, woke, woke agenda, wokeness, DEI, critical race theory, trans, gender identity, transgender, trans sports, and gender affirming care + children.

**Table 1. Issue Areas (Topics) and respective keywords**

| Issue Area | **Immigration** | **Inflation** | **Identity politics** | **Democracy** | **Public health** |
|---|---|---|---|---|---|
| **Keywords** | **Immigration** <br><br>**Illegal immigration** <br><br>**Illegal immigrants** <br><br>**Migrant crime** <br><br>**Border crisis** <br><br>**Border** <br><br>**Migrant** <br><br>**Immigrant** | **Cost of living** <br><br>**High price of food** <br><br>**High price of gas** <br><br>**Inflation** | **Identity politics** <br><br>**Woke** <br><br>**Woke agenda** <br><br>**Wokeness** <br><br>**DEI** <br><br>**Critical race theory** <br><br>**Trans** <br><br>**Gender identity** <br><br>**Transgender +sports** <br><br>**Transgender** <br><br>**Gender affirming care + children** | **Democracy** <br><br>**January 6** <br><br>**J6** <br><br>**Election denial** | **MAHA** <br><br>**Make America Healthy Again** <br><br>**RFK** <br><br>**RFK Jr.** <br><br>**Vaccines** <br><br>**Vaccination** <br><br>**Covid** <br><br>**Public health** |

We used the NLTK Python package [7], a package used for natural language processing (NLP) tasks, to tokenize and change to lower case as well as remove stop words and non-alphabetical characters from the text. We converted the date into days based on its placement in the order, for instance, Day 0 for October 29, 2024, Day 1 for November 1, 2024, etc. We then grouped by day and channel and converted the comments to a single string for easier matching. Each relevant keyword match would trigger a count under that keyword's associated issue, resulting in an issue frequency analysis across the days.

We then employed a clustering approach to analyze the thematic structure of the dataset. We used Sentence-BERT (SBERT), a sentence embedding transformer-based model for text embeddings [48]. SBERT creates numerical representations that capture the meaning of a sentence based on its context. We also used Uniform Manifold Approximation and Projection (UMAP) for dimensionality reduction [36], which projects high-dimensional embedding space into a form that preserves the local structure while being suitable for clustering. We also used Hierarchal Density-Based Spatial Clustering of Applications with Noise (HDBSCAN) for density-based clustering [11], which identifies dense groupings of semantically related comments without having to predefine the number of clusters. This method allowed us to identify coherent cluster topics within the comments. This pipeline has become increasingly common in computational social science for uncovering emergent issues in complex datasets [24, 25, 43].

Once the comments were clustered, we utilized GPT-4o, a large language model (LLM) [40], to generate labels for each cluster based on the contents of the comment for that cluster. LLMs like GPT-4o offer a flexible and efficient way to synthesize diverse inputs and produce coherent summaries or classifications. Hence, they are well-suited for labeling emergent issues in unsupervised settings [29]. We used Term Frequency-Inverse Document Frequency (TF-IDF) [51] to highlight relevant keywords in each comment. We then created a GPT-4o prompt (see Appendix A) that assigns a category based on the extracted keywords. The prompt also instructs GPT-4o to classify clusters according to our five predefined issue categories (see Table 1) while also permitting the generation of additional categories where necessary. This helped us eliminate comments that did not fall under any of the five predefined categories, ensuring our analysis remained focused on the primary issue areas.

Based on the proposed methodology, our goal is to demonstrate which topics or issue areas were most salient before the 2024 presidential election using issue frequency and clustering analysis.

4. Results

Our first analysis focused on the overall issue frequency based on occurrences of predefined keywords. As shown in Figure 2, *immigration* was the most frequently discussed topic, significantly outpacing the others. After immigration, *identity politics* also appeared frequently, while democracy, public health-related issues and inflation were mentioned with less frequency. These frequency differences among topics were statistically significant (Chi-square=99.85, df=4, p-value<0.0000001).

To further understand how these topics evolved over time, we looked at their keyword matches distributed across a seven-day period prior to and including the day of the election. Figure 3 shows the daily distribution of issue frequencies, showing that discussions on immigration remained consistently high across the period. Identity politics had sporadic peaks, particularly on days 3, 5 and 7. Democracy, public health-related issues and inflation remained stable with minimal fluctuations throughout the period.

**Figure 2. Frequency of issue areas identified by pre-defined keywords**

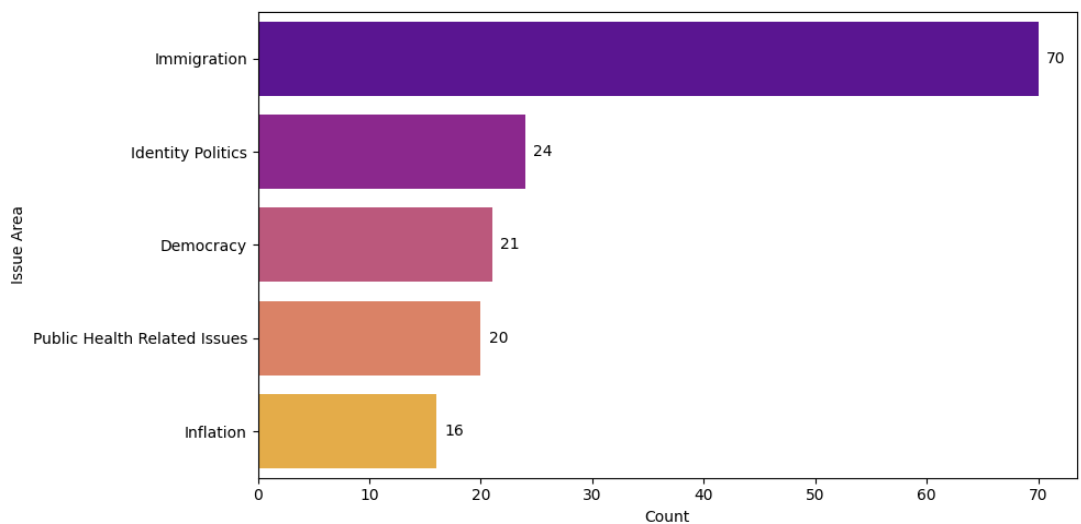

**Figure 3. Frequency distribution of issue areas identified by pre-defined keywords, by day (WSJ & NYT combined)**

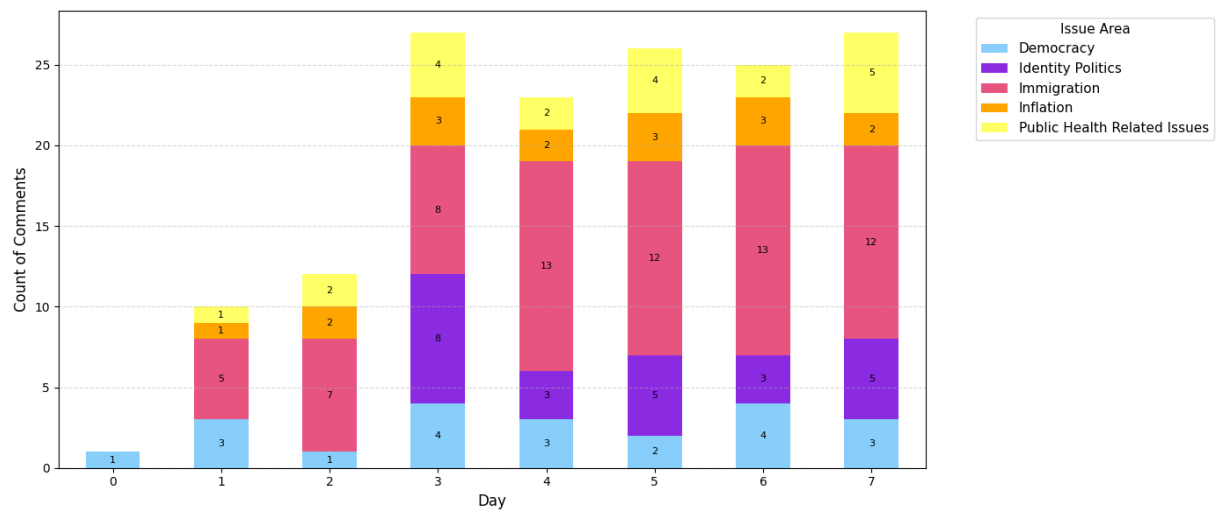

The results from our clustering analysis revealed distinct issue groupings in the dataset. Figure 4 shows a scatterplot of the clustered issues. Each point represents a single comment, the colors represent the assigned issue clusters, and the shapes indicate whether they were from NYT or WSJ videos. The plot shows that while some clusters are relatively well-formed, there is a noticeable overlap between issues, especially between democracy and immigration and identity politics with public health-related issues. While distinct issues can be identified, their boundaries are not sharply defined. This could be because discussions on democracy and immigration might frequently co-occur, making it difficult to separate them based on embeddings. It could also be because some user comments may discuss multiple topics leading to embeddings that intersect between clusters. Since UMAP was used to reduce high dimensional embeddings such as the ones SBERT creates – 5 dimensions down to 2, it can also cause artificial overlap. Despite this, our cluster analysis differentiated major issues and the separation between them enabled us to observe meaningful patterns in the clusters.

We can see that because WSJ and NYT occupy largely separate regions in the UMAP space, it indicates that the discourse between these two channels are semantically distinct. Meaning that although they may be discussing the same topics, the way they are talking about it is completely different. The language, framing, or narrative emphasis differs significantly between the two groups. We can see that for Democracy, which is spread across both, however for NYT it is denser. Identity Politics and Immigration on the other hand is more concentrated for WSJ, which can indicate a stronger focus on those topics and a narrower set of arguments or narrative on these issues whereas NYT is less dense indicating a broader diversity of views or more varied vocabulary. Inflation and Public Health are mixed, but they do cluster separately reinforcing the idea that although the topics overlap, the style and framing are different.

**Figure 4. Scatter plot of issue areas identified by cluster analysis**

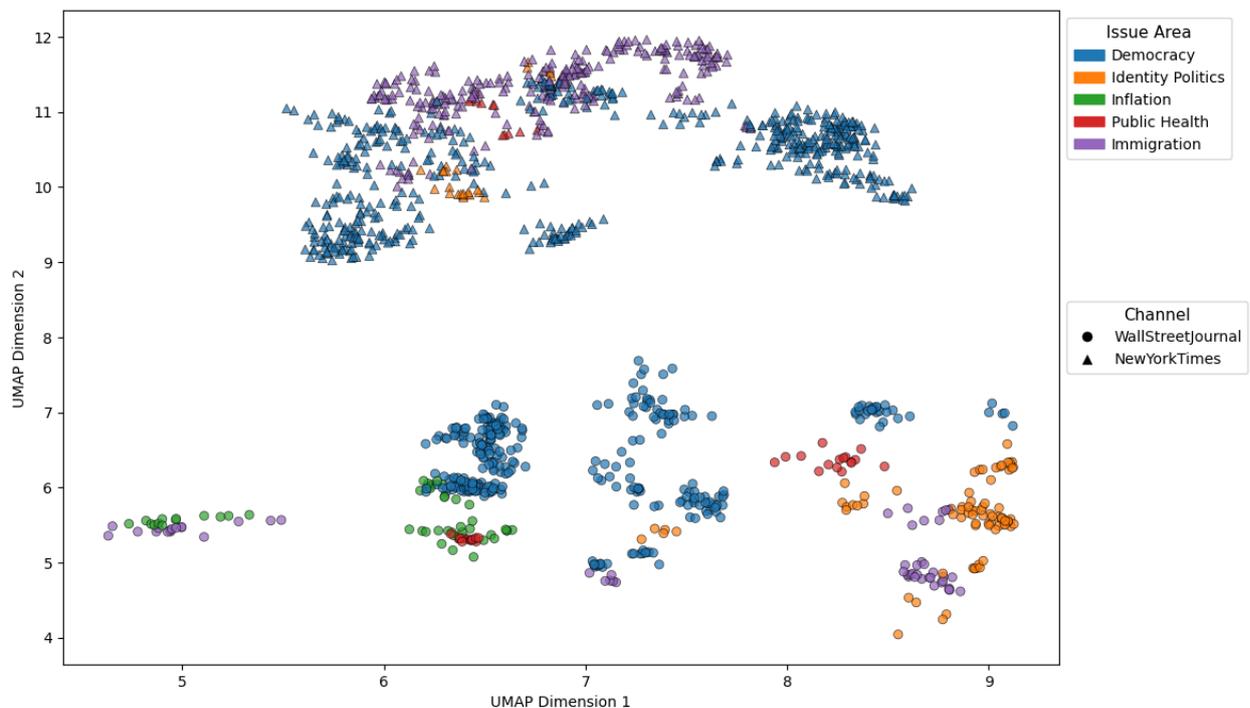

Figure 5 shows the distribution of issue clusters by channel. This shows that the NYT had much larger discussions surrounding democracy and immigration whereas the WSJ had a more even distribution of topics with significantly higher proportions of discussions on identity politics. Figure 6 shows the frequency of issue clusters, with democracy emerging as the most frequent, followed by immigration and then identity politics. Inflation and public health had lower counts. Overall, these differences in cluster frequency differences were statistically significant. (Chi-square=1719.26, df=4, p-value<0.0000001).

**Figure 5. Frequency of issue areas identified by cluster analysis, by YouTube channel (NYT & WSJ)**

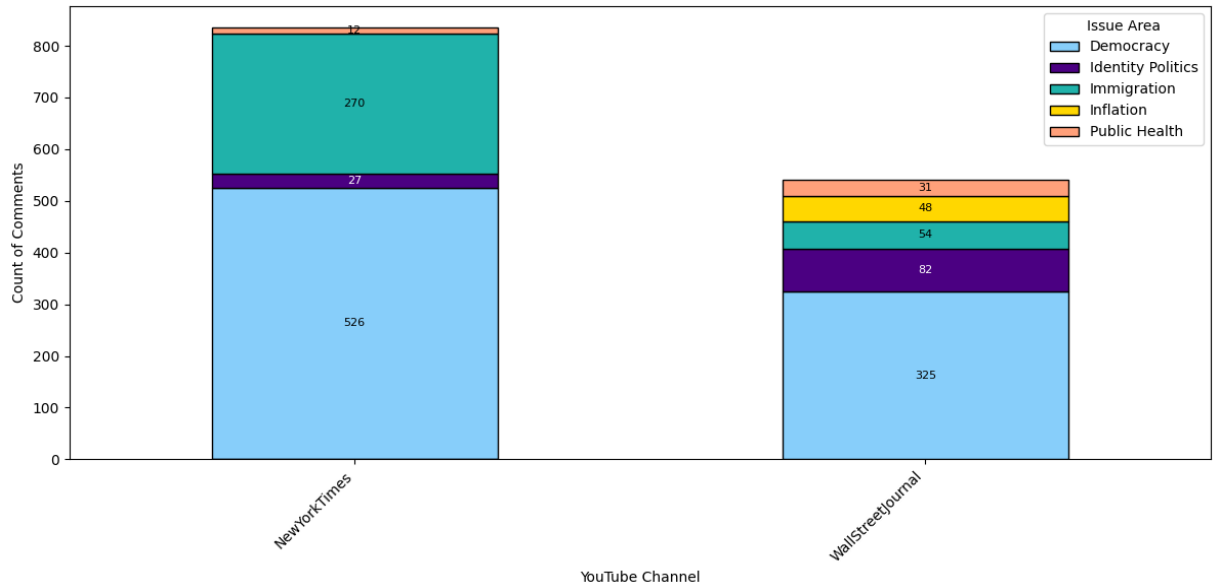

**Figure 6. Frequency of issue areas identified by cluster analysis, with NYT and WSJ channels combined**

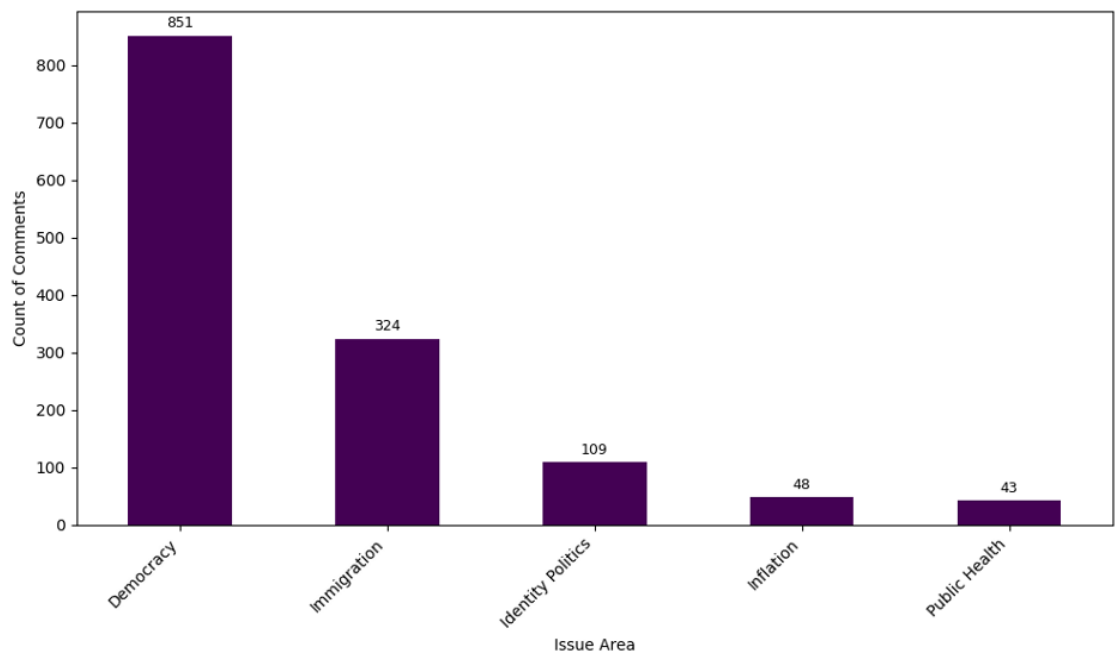

Figure 7 shows the distribution of issue clusters across the seven-day period. These results indicated that democracy and immigration grew steadily over time with its highest peak on the day before the election and second highest on the day of the election. The other remaining categories remained relatively stable, with slight increases of identity politics on days 3 and 7.

When we compare the results of the keyword matching and clustering analysis, we see that there is a strong alignment in immigration, democracy and identity politics being the dominant themes. Identity politics is the second most frequent topic in keyword matching, making it slightly higher than democracy, but it dropped to third place in the clustering analysis, with a much smaller representation compared to democracy and immigration.

**Figure 7. Frequency distribution of issue areas identified by cluster analysis, by day (WSJ & NYT combined)**

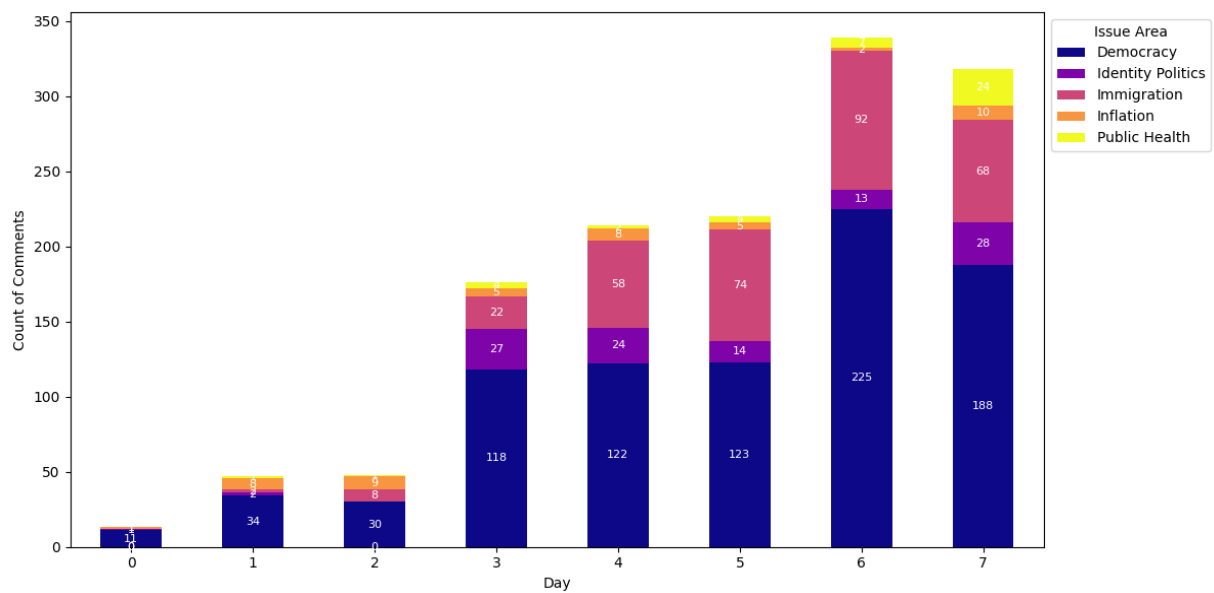

5. Discussion

Overall, our study demonstrates that inflation was not a major topic of discussion among the online commentators analyzed, compared to other issue areas. Depending on the method of analysis used – clustering or quantification of keyword frequency – immigration or democracy was the main issue being discussed by commentators of the election-related

YouTube videos analyzed. Identity politics was also a consistently more salient issue in these discussions than inflation, while public health was nearly on par with inflation as the least salient issue.

We began our analysis by categorizing key issue areas according to whether they were economic or non-economic and/or cultural/identity-related. In line with previous observation and political science scholarship on evolving value preferences in the industrialized world, this categorization informed our hypothesis. We posited that cultural or identity-related issues, such as identity politics and "wokeness," would be at least as salient, or even more salient, in potential voters' online election-related discussion than economic issues like inflation or the distinct issue of immigration. We tested this hypothesis using the methodology detailed above, and we can glean two key findings from the results. Firstly, our hypothesis was partially true. Though immigration was the most frequent issue area referenced in comments and the second most frequent cluster, inflation did turn out to be consistently less referenced in online discussions than most of the other issues, all of which are not of strictly economic nature. Secondly, both methods employed – scraping of online comments with NLP for predefined keyword searches and cluster analysis without predefined issue clusters – generated similar results, which demonstrates the relative and empirical criterion validity of the study, since neither method is a gold standard for the classification of the issues.

Regarding the first key finding, immigration, a cross-cutting issue that straddles both economic and cultural concerns, was significantly more salient than all other issues in the keyword distribution (see Figure 2). It was second to the wide-ranging cluster of democracy in the clustering analysis, but clusters around immigration remained far more salient than those around identity politics, inflation and public health (see Figure 6). In addition, in line with our hypothesis, both data analysis methods demonstrated that the most illustrative cultural issue area of identity politics was significantly more salient than inflation in online discussions. Finally, we can see in Figure 2 that public health-related issues were also slightly more frequently invoked than inflation, while in Figure 6 they were nearly on par with inflation as a salient issue in the clustering analysis without predefined keywords. The variation in the salience of identity politics could be because while the issue is frequently mentioned in keyword matching, its discussions may have been more dispersed across multiple topics, making it harder for the clustering analysis to identify it as a distinct, large group. This could mean that it was often discussed in broader conversations, potentially overlapping with other topics. Additionally, the unusually large salience of democracy as an issue area in the results of the cluster analysis could be due to the clustering tool identifying a broad range of sub-topics. The sub-clusters identified within the wide-ranging cluster of "democracy" include such disparate topics as oligarchic power, voting and inequalities. Nevertheless, both the frequency analysis of predefined keywords categorized into five issue areas and the clustering analysis demonstrate that democracy was more salient for voters than inflation.

Most interestingly in the clustering analysis, identity politics only emerged as a distinct cluster in the sample of comments starting on day 3 of the week preceding the election and continued to be more frequent than inflation up until election day (see Figure 7). In the keyword matching, identity politics emerged as an issue on Day 1 but also saw a significant leap in salience starting on Day 3 (see Figure 3). We see similar results regarding public health issues, around which we see more clusters forming, in comparison with inflation, between day 5 and day 7 (see Figure 7). Additionally, we can detect potential partisan differences in the issues driving voters. In Figure 5, we see that the largest clusters under the NYT channel were formed around the issues of democracy and immigration, whereas the WSJ channel saw the largest clusters around the issue of identity politics. This distinction between the issues identified by channel could indicate that, for example, the potentially right-leaning audience of the WSJ attributed greater importance to identity politics than the potentially left-leaning audience of the NYT.

These findings can have several sociopolitical implications. It confirms both the conclusions of post-election studies and our hypothesis that identity politics was a key election issue, well above inflation. It may also indicate that identity politics was specifically more salient for potential late-deciding voters and for right-leaning voters. Abundant literature shows the importance of this cohort of late deciders [6, 10, 21, 33]. Box-Steffensmeier et al. [9] summarize several conclusions of this literature by affirming that late-deciding voters "tend to be less partisan, less informed, and more ambivalent than early deciders" [9, p. 183]. This means that such voters are more susceptible to "short-term forces," which tend to influence vote choice and include positions on issues, since "salient issues change from election to election" [9, p. 181]. Our findings, showing that identity politics and public health related issues emerged more frequently closer to the election, indicate that these issues were potentially decisive in the choice of late deciding voters.

The research and conclusions drawn harbor limitations, most notably the lack of a gold standard to classify internet comments into key issue areas identified as relevant for presidential election sentiment in the specific context of 2024. There is inherent subjectivity to the definition of issue areas and the corresponding keywords we examined as potentially salient for voters, and the list of issues is clearly non-exhaustive. For example, we did not include other potential topics that were pertinent to the national and international context of 2024: the wars in Ukraine and Gaza, climate change and the environment, abortion rights, the controversy regarding President Biden's health, and the "economy" in its larger connotation, among others. However, the fact that two different methods of classification – keyword matching and clustering analysis – provided somewhat similar results is reassuring that the key research questions are being answered with some degree of validity. Additionally, the subjectivity in selecting the issue areas is diminished by the extensive literature review implemented to generate the five key issues. Though other topics could have been included, we decided to limit them to five overarching issue areas that were most representative of the issues supposedly driving the election, according to the press, survey research and extant political science research.

Additionally, there are limits to what can be inferred or interpreted from the frequency of keywords mined in YouTube users' comments. As N. Beauchamp [4] highlights in his analysis of Twitter textual data as a predictor of opinion polls, similar methods have been critiqued in the past. He points especially to the study by Tumasjan et al. [55], which conducted content analysis of tweets and concluded that "the mere number of messages mentioning a party reflects the election result" [55, p. 178]. This was a pioneering study in the early days of Twitter and widespread social media use, which was meant to provide initial evidence for the importance of Twitter as "a platform for political deliberation" [55, p. 183] and to demonstrate the utility of opinion mining in social media as a "valid indicator of political sentiment" [55, p. 178]. Nevertheless, the study was critiqued for being reductive and nonrepresentative, particularly in its choice of parties selected in the study [31]. As Jungherr et al. [31] note, if the Pirate Party had been included in the study by Tumasjan et al. [55], it would have incorrectly been predicted to be the election winner.

We acknowledge that our study could have a similar limitation, especially regarding the method of quantifying the frequency of keywords in user comments as an indicator of issue salience. However, we significantly temper this limitation by cross-verifying the keyword frequency with the AI-driven cluster analysis method, which does not use predefined clusters as a starting point. As we have shown, the results of both methods were only slightly different, given the breadth of the sub-topics clustered under the issue of "democracy." Besides that divergent result, the cluster analysis resulted in a similar frequency distribution of the issue areas, with immigration and identity politics still more frequent than inflation among the salient topics. This not only further partially validated our hypothesis that inflation was not the primary issue driving vote choice, but it illustrates the relative validity of the opinion mining methods used.

We did not use standard topic modeling methods like Latent Dirichlet Allocation (LDA) because they do not work well with informal text such as user comments [26, 58]. LDA relies on word co-occurrence and assumes fixed topic distributions, which often results in incoherent topics when applied to sparse, nuanced content [17, 42]. Although BERTopic [22] offers improvements by combining embeddings with clustering and keyword extraction, we opted for a custom pipeline to gain more flexibility. BERTopic's default labeling approach (which is TF-IDF) does not easily adapt to predefined analytical categories or allow for prompt-based refinement. By using SBERT, UMAP, HDBSCAN, and GPT-4o, we retained the strengths of BERTopic's structure while enabling domain-specific, interpretable labeling tailored to our research objectives [24, 25].

A possible next step in our study could be to perform sentiment analysis of the same sample of comments. If there is anything close to a gold standard in opinion mining, it may involve some level of sentiment analysis. Indeed, R. Liu et al. [35] used Twitter sentiment analysis, through NLP, in place of polling data to analyze whether positive or negative sentiment attributed by Twitter users to Hillary Clinton and Donald Trump could forecast the 2016 presidential election results. The authors note that such methods have been increasingly popular in the field of computer science since the expansion of social media and studies using Twitter data to illustrate their utility in predicting election results have proliferated since the 2010s. Many studies use sentiment analysis methods to forecast election results based on positive or negative sentiment towards individual political leaders [12] or candidates [57], or political parties [44]. Alternatively, Swamy et al. [53] use opinion mining methods to assess to which extent social media users' beliefs on who will win elections can forecast the eventual winner. We maintain that our study distinguishes itself from these previous studies in its focus on *issues* rather than individuals. Applying sentiment analysis to issues instead of nominal individuals or parties may require further research and methodological refinement, but it is a potential avenue for future applications of the methodology described in this paper.

Given our findings, we propose a future line of research that can improve the methodology deployed and confirm our general findings. We could measure how often these issues emerge in comments across time, from December 1, 2020 (first day of first month after last presidential election) until the November 5, 2024 election, to identify peaks. We could then subsequently cross-reference this data with election results by state, to identify which issues were more salient in which states and even counties. In addition, we could also cross-reference this data with the known proportional representation of registered voters (Democrats, Republicans and Independents) by geolocation in the United States (e.g., state, county, Zip code or even census tract), provided that the user profiles are public. This would provide novel data for inferring which issues contributed to the way the election swung, particularly in the most contested states of Michigan, Wisconsin, Pennsylvania, etc.

**Conclusion**

Our paper innovates the political prediction field by employing novel methods based on AI to glean insight into the issues potential voters were discussing online regarding the 2024 election. Two specific methods were used to analyze thousands of user comments on election-related YouTube videos on the channels of the New York Times and the Wall Street Journal: (1) NLP to identify the frequency of five predefined issues – immigration, identity politics, inflation, democracy and public health – based on the count of predefined keywords corresponding to each issue; (2) clustering analysis to identify clusters of topics that were *not* predefined, then ChatGPT to label the identified clusters according to our five predefined

23issue areas. This analysis allowed for gauging the issues that were most salient among potential voters, providing insight into the key issues driving voter choice in 2024.

In our analyses, immigration and democracy appear as the key electoral issues among potential voters, depending on the method employed to analyze the data. However, both methods showed that identity politics were high in frequency compared to inflation. More specifically, the NLP-driven method for quantifying the relative frequency of the five issues showed that immigration was significantly more salient than all other issues, while identity politics was the second most salient issue above democracy, public health and lastly inflation. The clustering analysis provided only slightly different results, in that democracy appeared as the largest cluster, followed by immigration, identity politics, inflation and lastly public health. Overall, regarding our hypothesis, inflation was consistently less salient than immigration, democracy, and identity politics regardless of the method used to analyze user comment data.

**Data Availability Statement**

Supporting materials such as data and code for this article is available at Simoes et al. (2025) at https://doi.org/10.7910/DVN/8C7VCQ.

**Funding:** This work was supported by the National Science Foundation and the U.S. Office of Personnel Management Cybercorps (Grant No. 1946619).**References:**

1. Abeysinghe, B., Vulupala, G. R., Bourgeois, A. G., & Sunderraman, R. (2022). Unsupervised User Stance Detection on Tweets Against Web Articles Using Sentence Transformers. *2022 IEEE International Parallel and Distributed Processing Symposium Workshops (IPDPSW)*, 1163–1169. https://doi.org/10.1109/IPDPSW55747.2022.00190

2. Abramowitz, A. I., & Webster, S. (2016). The rise of negative partisanship and the nationalization of U.S. elections in the 21st century. *Electoral Studies*, *41*, 12–22. https://doi.org/10.1016/j.electstud.2015.11.001